\documentstyle[12pt,cite]{article}
\begin{document}

\begin{center}
{\large {\bf DEEP INELASTIC LEPTON SCATTERING IN  \\

\vspace{0.4cm}

NUCLEI AT $ x > 1$ AND THE NUCLEON 

\vspace{0.5cm}

SPECTRAL FUNCTION}}
\end{center}

\vspace{1cm}

\begin{center}
P. Fern\'andez de C\'ordoba, E. Marco, H. M\"{u}ther, E. Oset
and Amand Faessler
\end{center}

\vspace{1cm}

\noindent
{\it Institut f\"{u}r Theoretische Physik, Universit\"{a}t T\"{u}bingen,
72076 T\"{u}bingen, Germany.}

\vspace{3cm}

\begin{abstract}
{\small{The nuclear structure function $F_{2A} (x)$ has been studied in the 
Bjorken limit for $(l, l')$ scattering on nuclei in the region
of $x > 1$ and was found to be very sensitive to the information
contained in the nucleon spectral function in nuclei, particularly the
correlations between momenta and energies in the region of large momenta.
Calculations were done in a local density approximation using two
different spectral functions for nuclear matter. Results are compared
to those obtained for a spectral function which has been evaluated
directly for the finite nucleus, $^{16}O$, under consideration.
For values of $x$ around 1.5 and larger the quasiparticle contribution 
is negligible, thus stressing the sensitivity of the present
reaction to the dynamical properties of nuclei beyond the
shell model approach. Several approximations which are usually employed 
in studies of the EMC effect have been analyzed and their inaccuracy in
this region is demonstrated.
The results stress the fact that the nuclear structure
function contains important infor\-ma\-tion on 
nuclear dyna\-mi\-cal correlations. Therefore further
measurements of $F_{2A} (x)$ in that region and for many
nuclei would be most welcome.}}
\end{abstract}

\newpage

\section{Introduction}

Deep inelastic scattering of leptons on nuclei in the region of asymptotic
freedom 
and the ratio $R (x)$ of the nuclear structure function compared to the
corresponding one for the deuteron
(EMC effect \cite{1}) has been one of the topics in the
interface of nuclear and particle physics intensively studied in the recent
past \cite{2,3,4,5}. After multiple discussions, pionic effects
\cite{4,6,7,7A}, binding effects \cite{8,9,10} and Fermi motion \cite{3,6,11}
have turned out to be important ingredients to describe the main
characteristics of the
ratio $R (x)$: the small enhancement beyond unity around $x \simeq 0.1$, the
small depletion around $x \simeq 0.6$ and the steady increase around
$x \simeq 0.8$ and above, respectively.

The increase of $R (x)$ at $x$ close to one was soon identified to be
a consequence of Fermi motion \cite{3,6,11} and this is one
of the points where there seems to be consensus among scientists. Probably
the lack of controversy on this issue prevented 
a systematic exploration of the region of
 $x >1$,
in spite of the fact that in nuclei only the variable  $x_{A} =
- q^2 /2M_A q^0$ is limited between 0 and 1, while
$x = - q^2 / 2 M q^0$ varies from 0 to $M_A /M$. Since 
$F_{2N} (x)$, the structure function of  a free nucleon,
 is zero for $x > 1$, the fact that $F_{2A} (x)$ is different from zero 
for $x > 1$ must necessarily be attributed to modifications of the nucleon
properties inside nuclei, i.e.~to genuine many body 
effects. The region of $x > 1$ is hence a source of very interesting
information on 
nuclear properties, as we shall see.

The fact that Fermi motion was so important at $x \simeq 1$ induced
people to investigate the region of $x > 1$ with the same idea
\cite{10,11,12,13}.
These works essentially employed phenomenological nuclear
matter momentum distributions, with the main conclusion that a 
nonvanishing value of $F_{2A} (x)$ for $x > 1 + k_{F}/M$ (with $k_{F}$
the Fermi momentum) would require a tail in the momentum distribution
$n (\vec{k})$ for $k > k_{F}$.
The occupation number $n (\vec{k})$ is different from zero
for  $k> \, k_{F}$ as soon as the effects of a residual NN interaction
are considered leading to a correlated many-body system of Fermions.

However, it is quite dangerous to reduce the effects of correlations to
a discussion of a momentum distribution only,
because in a system of interacting Fermions the energy and
momentum distributions are correlated by means of the spectral function,
and the simultaneous consideration of both distributions is necessary
in principle, and also in the practice of the present case, as we shall see. 
Disregarding the energy and momentum correlations leads sometimes to quite
erroneous results, like in the study of the $\Lambda$ mesonic decay in
nuclei where the results for the width based exclusively on the nucleon
momentum distribution in nuclei are three orders of magnitude bigger
than the results obtained with a spectral function, or the 
experimental results \cite{14}. The same warning, in the present context,
was raised in ref. \cite{15} where $F_{2A} (x)$ was evaluated for
$^3 He$. Further work in this direction was done in refs.
\cite{15A,15B,15C}.

In the present work we have evaluated $F_{2A} (x)$ for several nuclei by
using spectral functions of infinite nuclear matter and the local
density approximation. This approximation is good when dealing
with volume processes like the present one. In order to
quantify uncertainties from the many body approach used, we have performed
the calculations using two spectral functions evaluated with rather different
methods. On the other hand, since large momentum components are
necessarily involved in the process, relativistic corrections are bound
to be relevant \cite{16,17} and we have worked within a relativistic
approach.

We evaluate $F_{2A} (x)$ for values of $1 \leq x \leq 1.5$. At $x= 1.5$ the
value of the structure function decreases
by about three orders of magnitude with
respect to $x = 1$, but is still well within measurable range. We also
show results using the momentum distributions alone, within several
approximations, and find appreciable differences (of two or more
orders of magnitude) with respect to the accurate results using the
spectral functions. These results show that the study of nuclear
structure functions at values of $x >1$ is a very interesting
tool to learn about nuclear dynamical correlations beyond 
the nuclear properties 
described in a shell model approach.

In this paper we not only discuss results obtained within  the local
density approximation but also consider a spectral function calculated
directly for the nucleus $^{16}$O. After this introduction we
discuss in section 2 the calculation of the nuclear structure function
and its relation to the spectral function. The various approaches
for the spectral functions are presented in section 3. The
results of our studies are presented and discussed in section 4. The
last section summarizes the main conclusions.

\section{The nuclear structure function}

Deep inelastic electron (or muon) scattering on an unpolarized nucleon
can be described in terms of
two structure functions, $W_1 (x, Q^2) \, , \, W_2 (x, Q^2)$, where the
Bjorken variable $x$ is given by

\begin{equation}
x = \frac{-q^2}{2 pq} = \frac{Q^2}{2 pq}
\end{equation}

\noindent 
with $q$ the momentum of the virtual photon
and $p$ the momentum of the nucleon. In the Bjorken limit, 
$q^0 \rightarrow \infty, \, Q^2 \rightarrow \infty$ and $x$ fixed, it
is common to define the structure functions $F_1$ and $F_2$ which 
depend only on the variable $x$, up to some smooth logarithmic dependence
on $Q^2$ from QCD corrections. In this limit one has

\begin{equation}
\begin{array}{ll}
\frac{p . q}{M} W_2 (x, Q^2) & \equiv F_2 (x)\\[2ex]
M W_1 (x, Q^2) & \equiv F_1 (x)
\end{array}
\end{equation}

\noindent
and $F_2 (x) \, , \, F_1 (x)$ are related by the Callan-Gross relation

\begin{equation}
2x F_1 (x) = F_2 (x)
\end{equation}

Using these structure functions $W_1$ and $ W_2$ the hadronic tensor
for the absorption of the virtual photon can be written:

$$
W'^{\mu \nu} = ( - g^{ \mu \nu} + \frac{q^\mu q^\nu}{q^2}) W_1 + 
p'^{\mu} p'^{\nu} \frac{W_2}{M^2} 
$$

\begin{equation}
p'^{\mu} = p^{\mu} - \frac{p.q}{q^2} q^{\mu}
\end{equation}

It is practical to work in a frame where $\vec{q}$ is parallel to the
$z$ direction. Adopting this frame and inspecting the transversal
$W'^{xx}$ component in the Bjorken limit, one finds that 
the term proportional to $W_2$ in eq. (4) vanishes and $W'^{xx}$ is
related to $W_{1}$ for nucleons in nuclei with the same coefficient in
front as for one nucleon in the vacuum, independent on its momentum or
energy.
This allows one to write the structure function $F_{1A}$ derived from
lepton scattering from a nucleus with baryon number $A$ in a 
nonrelativistic formalism using eq. (2).

\begin{equation}
\frac{F_{1A} (x_A)}{M_A} = \int \frac{d^3 p}{(2 \pi)^3}
\int_{- \infty}^{\mu} d \omega \, {\cal S}_h (\omega,p) \frac{F_{1N} (x_N)}{M}
\theta (x_N) \, \theta (1 - x_{N})
\end{equation}

\noindent
where ${\cal S}_{h}(\omega,p)$ denotes the hole spectral function, i.e. the
probability of finding a nucleon with energy $\omega$ and momentum $p$
in the nucleus, the integration limit $\mu$ the chemical potential or
Fermi energy and

\begin{equation}
x_{A} = \frac{- q^2}{2 p_A q} = \frac{- q^2}{2 M_A  q^0} \, ;\quad
x_N = \frac{- q^2}{2pq} \, ; \quad p \equiv (\omega, \vec{p})
\end{equation}

\noindent
Instead of $x_A$ one normally uses  the variable $x$,

\begin{equation}
x = \frac{ - q^2}{2 M q^0} = \frac{M_A}{M} x_A
\end{equation}

\noindent
so we will write $F_{1 A}$ and $F_{2 A}$ as functions of $x$ from now on.

By means of eq. (3) for nuclear targets we can calculate $F_{2 A}$ which is
the structure function used in studies of the $EMC$ effect. We then write

\begin{equation}
F_{2 A} (x) = \int \frac{d^3 p}{(2 \pi)^3} \int_{- \infty}^\mu
{\cal S}_h (\omega, p) \frac{x}{x_N} F_{2N} (x_{N}) \theta (x_N) 
\theta ( 1 - x_N)
\end{equation}

Instead of using the spectral function ${\cal S}_{h}$ calculated
directly for the nucleus under consideration, it is common practice to
employ a local density approximation and represent this spectral function
in terms of a spectral function $S_{h}(\omega,p;\rho)$ 
evaluated for infinite nuclear matter at various densities $\rho$ which
is normalized by

\begin{equation}
4 \int \frac{d^3 p}{(2 \pi)^3} \int^{\mu}_{- \infty}
d \omega \; S_h (\omega, p;\rho) = \rho\, ,
\end{equation}

\noindent
with a factor 4 on the left side of this equation to account for the
spin-isospin degeneracy of symmetric nuclear matter. Assuming a density
profile $\rho(r)$ for the finite nucleus to be studied, one can
determine the local density approximation for the spectral function of
this nucleus by

\begin{equation}
{\cal S}_{h} (\omega, p) = 4 \int d^3r S_h \left(\omega,
p;\rho(r)\right) \,
\end{equation}

\noindent
which ensures that

\begin{equation}
4 \int d^3 r \int \frac{d^3 p}{(2 \pi)^3} \int^{\mu}_{- \infty}
d \omega \; S_h \left(\omega, p;\rho(r)\right) = \int \frac{d^3 p}{(2 \pi)^3}
\int^{\mu}_{- \infty} d \omega \;{\cal S}_{h}(\omega, p) = A\; .
\end{equation}

For nuclear matter the spectral function can be evaluated in terms
of the nucleon selfenergy $\Sigma (\omega, p)$ by

\begin{equation}
S_h (\omega, p) = \frac{1}{\pi} \frac{Im \Sigma (\omega, p)}{[\omega
 - \varepsilon 
(\vec{p}) - Re \Sigma (\omega, p)]^2 + [Im \Sigma (\omega, p)]^2}
\end{equation}

\noindent
where we have dropped the variable $\rho$ identifying the density
dependence of the self-energy and spectral function. In eq. (12),
$\varepsilon (\vec{p})$ is used to represent the nucleon kinetic energy.

Relativistic corrections accounting for the kinematics of the nucleon have
been included in deep inelastic scattering.
Prescriptions based on the normalization of
the relativistic current operator lead to corrections
of the static, or shell model, structure function of the nucleus
\cite{18}. Further corrections have been considered in \cite{16,17} at
$x \simeq 1$. In ref. \cite{19} a different relativistic treatment is developed 
which allows to write all  quantities in terms of the nucleon
propagators. Only the region $ 0 < x < 1$, which was measured by the
$EMC$ collaboration, is studied there. As in ref.
\cite{4,6,7}, pionic corrections are shown to be relevant in the 
region of  $x < 0.6$, but they play no role in the region  $x > 1$ which
we study here.

Employing the treatment described in ref.\cite{19}, which uses
a relativistic spectral function from the beginning, one can avoid 
introducing any flux factors as  in ref.\cite{18} to
account for relativistic corrections in non-relativistic nuclear wave
functions. 

Since the relativistic corrections are important here we take advantage
to discuss briefly and complement the details of ref. \cite{19}.

In fig. 1a we show the Feynman diagram that symbolizes the deep inelastic
process on a nucleon. The final hadronic state $X$ will contain at least one
baryon and will have baryonic number one. In a nucleus the nucleon $N$ will 
have a certain momentum and energy distribution given by the spectral
function. The most practical way to take this into account and have a 
covariant formulation  of the nuclear problem is to fold the amplitude
 in fig. 1a and convert it into a many body diagram for the selfenergy of an
electron in the nuclear medium, fig. 1b. Here the nucleon $N$ in
fig. 1a gets converted into a hole line  and with the
baryon existing in $X$ it completes a fermionic loop. In section 3 of 
ref. \cite{19} one evaluates this selfenergy in infinite
nuclear matter and, by means of the local density approximation, the $(e,e')$
cross section, is related to the imaginary part of the electron
selfenergy. The imaginary part of the $e$ selfenergy is evaluated using 
Cutkosky rules and this means that the intermediate states $e'$ and
$X$ are placed on shell in the integrations over the momenta of these
states. The formalism is originally covariant in the sense that 
everything is written in terms of propagators of the particles and
we can write the nucleon propagator in a covariant relativistic way.
However, Cutkosky rules select the imaginary part of the propagator 
of nucleon $N$ for the occupied states and when doing that the
apparent covariant structure might not show up clearly.

In our formalism we start from a free nucleon propagator which we 
split into its positive and negative energy parts \cite{20}

\begin{equation}
\frac{p \!\!/  + M}{p^2 - M^2 + i \epsilon} \equiv \frac{M}{E (\vec{p})}
\left\{ \frac{\sum_r u_r (\vec{p}) \bar{u}_r (\vec{p})}{p^0
- E (\vec{p}) + i \epsilon} + \frac{ \sum_r v_r (- \vec{p}) \bar{v}_r
 (-\vec{p})}{p^0 + E (\vec{p}) - i \epsilon } \right\}
\end{equation}

\noindent
where $M, E (\vec{p}\,)$ are the nucleon mass and the relativistic nucleon
energy $(\vec{p}\,^2 +  M^2)^{1/2}$ and $u_r (\vec{p}) \, , \, v_r (\vec{p}\,)
$ are
the ordinary spinors which we take normalized as $\bar{u}_r (\vec{p}\,)
u_r (\vec{p}\,) = 1$. We recall that $u_r (\vec{r}\,)$ are functions of three
momentum and they are the only spinors which appear in our
framework.

In order to account for binding and momentum distribution of the occupied
nucleons we need the nucleon propagator in the nucleon medium.

Note that even if a nucleon is off shell,
$p^0 \neq E (\vec{p}\,)$,
in the propagator of eq. (13) and we have
$ p\!\!/  + M \rightarrow p^0 \gamma^0 - \vec{p} \vec{\gamma} + M$, the
positive energy part has the Dirac structure $2 M \sum_{r} u_r (\vec{p})
\vec{u}_r (\vec{p}) = E (\vec{p}) \gamma^0 - \vec{p} \vec{\gamma} + 
M$, corresponding to on shell nucleons of momentum $\vec{p}$.

Following a standard relativistic notation \cite{21} the nucleon
propagator in a spin saturated system  would be

\begin{equation}
G (p^0,p) = \frac{1}{p\!\!/ - M - \Sigma^s - \Sigma^v \gamma^0}
\end{equation}

\noindent
which includes a scalar and vector terms in the nucleon
selfenergy (the inclusion of a term of the type $\vec{\gamma} \vec{p}$
does not change the arguments and conclusions which follow). The 
extraction of hole and particle spectral functions requires the
evaluation of $Re \Sigma^{s,v}$ and particularly $Im \Sigma^{s,v}$,
which is a non trivial task.

We respect the structure of eq. (14) but follow a different path 
 in order to single out the imaginary part of the
positive energy piece of the nucleon propagator. We start from
the realization that for this latter purpose, in  a perturbative expansion
of the propagator of eq. (14) in  terms of the free
propagator of eq. (13), the terms of positive energy will be singular
and dominate over those of negative energy. This allows us to 
write the desired part of the propagator as

$$
\tilde{G} (p^0, p) = \frac{M}{E (\vec{p})} \sum_r u_r (\vec{p}) \bar{u}_r (\vec{p}) 
\frac{1}{p^0 - E (\vec{p})} + 
$$

$$
\frac{M}{E (\vec{p})} \sum_r \frac{u_r (\vec{p}) \bar{u}_r (\vec{p})}{p^0
- E (\vec{p})} \Sigma (p^0, p) \frac{M}{E (\vec{p})} \sum_s \frac{u_s
(\vec{p}) \bar{u}_s (\vec{p})}{p^0 - E (\vec{p})} + ...
$$

\begin{equation}
= \frac{M}{E (\vec{p})} \sum_r \frac{u_r (\vec{p}) \bar{u}_r (\vec{p})}
{p^0 - E (\vec{p}) - \bar{u}_r (\vec{p}) \Sigma (p^0,p) u_r (\vec{p})
\frac{M}{E (\vec{p})}}
\end{equation}

This expansion is rather useful because both $\Sigma^s $ and $\Sigma^v
\gamma^0 $( and $\vec{\gamma} \vec{p})$ are diagonal in the base
of the $u_r (\vec{p})$ spinors, which converts eq. (15) in an
ordinary geometric series, not a matricial series, which can be summed
trivially as shown  in the last step of eq. (15).

It might look surprising that one obtains a Dirac structure $u_r (\vec{p}\,)
\bar{u} (\vec{p}\,)$ in $\tilde{G} (p^0,p)$ as for 
free nucleons, even when the renormalized nucleons will be off shell.
This is less striking if one recalls that also in eq. (13) the positive
energy part (corresponding to $\tilde{G}$) has the same structure
$u_r (\vec{p}) \vec{u}_r (\vec{p})$ even if the nucleon is
off shell. In any  case the structure might not match the one
coming from eq. (14) and one has lost the covariance shown by
eq. (14). The reason for this loss of covariance is that one loses
terms with admixture of the positive and negative parts of the nucleon
propagator of eq. (13) in the perturbative exapansion. This is,
in the sum of eq. (15) one is summing terms of the type of fig. 2a,b,c,d,
where in the intermediate fermion lines one only has the part of positive
energy of the propagator. One is missing terms of the type of fig. 2e
(where the line pointing down stands for the negative energy part of 
the free nucleon propagator), which would naturally appear in a covariant
expansion of eq. (14).

On the other hand while all the terms of fig. 2c,d etc. are summed up 
automatically  in eq. (15) in terms of a selfenergy given exclusively
by the term in fig. 2b, the second order terms of fig. 2e is not
accounted for. We argued that this latter term (which has an intermediate
propagator of order 1/2M, and is of $\rho^2$ type) should be small with
respect to the diagrams contained in figs. 2b, c, d, ... But in any case it 
can be included as a nucleon selfenergy  part in the sum of eq. (15) and 
then diagrams e,f, etc. would be automatically included. This means that
even if the covariance is lost in eq. (15) one can still regain all
the terms in the series by including these mixed terms in the selfenergy 
$\Sigma$
appearing in eq. (15). Of course, this selfenergy is now different to the 
one appearing in eq. (14). These mixed terms are also diagonal in
$u_r (\vec{p})$ and do not change the structure of eq. (15). This is the
philosophy which we follow, only that the diagrams of fig. 2e, 2f are not 
evaluated, although they are implicitly accounted for
as we pass to discuss. The reason is that
these diagrams only contribute to the real part of
$\Sigma$, not to the imaginary part, and in our scheme, which 
evaluates accurately $Im \Sigma$, there are pieces missing in the real 
part of $\Sigma$ which are added phenomenologically in order to ensure  the
exact experimental binding energy of each nucleus \cite{19}. The
particular structure of eq. (15) allows one to write

$$
\tilde{G} (p^0, p) = \frac{M}{E (\vec{p})} \sum_r u_r (\vec{p}\,) \bar{u}_r
(\vec{p}\,) [\int^\mu_{- \infty} d \omega \frac{S_h (\omega,p)}{p^0
- \omega - i \eta}
$$

\begin{equation}
+ \int_\mu^\infty d \omega \frac{S_p (\omega, p)}{p^0 - \omega + i \eta}
]
\end{equation}

\noindent
with the relationships

$$
\begin{array}{rr}
S_h (p^0, p) & = \frac{\displaystyle{1}}{\displaystyle{\pi}
} \frac{\displaystyle{\frac{M}{E (\vec{p}\,)} Im
\Sigma (p^0,p)}}{\displaystyle{[p^0 - E (\vec{p}\,) - \frac{M}{E (\vec{p}\,)}
Re \Sigma (p^0, p)]^2 + [\frac{M}{E (\vec{p}\,)} Im \Sigma (p^0, p)]^2}}
  \\[2ex]
&  \hbox{for}\, \,  p^0 \leq \mu
\end{array}
$$

\begin{equation}
\begin{array}{rr}
S_h (p^0, p) & = - \frac{\displaystyle{1}}{\displaystyle{\pi}
} \frac{\displaystyle{\frac{M}{E (\vec{p}\,)} Im
\Sigma (p^0,p)}}{\displaystyle{[p^0 - E (\vec{p}\,) - \frac{M}{E (\vec{p}\,)}
Re \Sigma (p^0, p)]^2 + [\frac{M}{E (\vec{p}\,)} Im \Sigma (p^0, p)]^2}}
  \\[2ex]
&  \hbox{for}\, \,  p^0 > \mu
\end{array}
\end{equation}

\begin{equation}
 k_{F,p} (\vec{p}\,) = [3 \pi^2 \rho_p (\vec{p}\,)]^{1/3} \, \, \,
k_{F,n} (\vec{r}\,) = [3 \pi^2 \rho_n (\vec{r}\,) ]^{1/3}
\end{equation}

\noindent
where for simplicity $\Sigma$ is now $\bar{u} \Sigma u$
which is independent of spin.

By means of this new nucleon propagator the modifications introduced
by our relativistic formalism, described in detail in ref. \cite{19},
are rather intuitive, easy to employ and can be summarized as:

\begin{itemize}
\item[i)] The normalization of the spectral function which ensures the proper
normalizations of the charge (or baryonic charge) of a nucleus is
exactly the same  as in eq.(9).
 However, the spectral function of eq.(12) is now
replaced by eq. (17)

\item[ii)] On the other hand the structure function $F_{2A} (x)$ of
eq.(8)
is replaced by

\begin{equation}
F_{2A} (x) = 4 \int d^3 r \int \frac{d^3 p}{(2 \pi)^3} \frac{M}{E (\vec{p})}
\int_{- \infty}^{\mu} d \omega \, S_h (\omega,p) \frac{x}{x_N}
F_{2N} (x_N) \theta (x_N) \theta (1 - x_N)
\label{eq:strucr}
\end{equation}

\end{itemize}

\noindent
where the relativistic factor $\frac{M}{E (\vec{p}\,)}$ plays the role
of a Lorentz contraction factor, appearing in the probability per unit time
of electron collision with the nucleon, and which remains in the formula
of the nuclear cross section because one divides the sum of all
probabilities by a unique electron flux, the one of the electron with 
respect to the CM of the nucleus.

The questions of normalization and conservation of baryonic number,
which have been the subject of much attention \cite{2}, are discussed 
in detail in \cite{19}.

There are other terms which would be included in a covariant formalism
of the $(e, e')$ reaction and do not appear in our formalism. These are terms which
have a negative energy state coupled to the hadronization vertex,
as shown in fig. 3. Once again  such terms are reduced by the large 
energy denominator of the negative energy state. 
Only in cases when
one uses an
 operator which magnifies the $N \bar{N}$ coupling with
respect to the $NN$, as in the case of the
axial charge, such terms can be relevant \cite{22}, although more accurate
nonperturbative calculations find smaller results  \cite{23}. With 
the use of scalar potentials
$\Sigma^s$  smaller than the typical ones in the Walecka model
when one imposes constraints from information in the negative energy
sector, as done in \cite{24},  terms like in fig. 3
would be of the order of 10-15 $\%$ if one has an operator like in the
axial charge, $\gamma^0 \gamma_5$, but much smaller than this if
electromagnetic current operators are used \cite{23}.

Furthermore, in as much as one assumes that the structure functions for the
positive or negative energy states are the same and one uses a nucleon spectral
function which conserves the baryonic number, 
 one would be including the strength of these 
pieces into the scheme   which we follow, up to small differences 
coming from different medium corrections to the positive and
negative energy states. Estimates based on the findings of \cite{23} would
put the difference between this covariant scheme and ours at the level of 
1-2$\%$ in the EMC region and probably a few percent in the $x >1$ region
that we explore.

In practical terms our scheme 
respects special relativity in the positive energy
sector and amounts to using free $u_r (\vec{p}\,)$ spinor in the
evaluation of the matrix elements of the $\gamma^* N \rightarrow
 X$ process while
keeping the proper energy and momentum balance in the $\delta$ function
of conservation of fourmomentum, with the $\omega, p$ distribution of the
occupied nucleon given by the spectral function and the energy  and
momentum of the final states in $X$ being those of their asymptotic
states. This is in fact the most standard method in the study of
many nuclear processes involving scattering or decay. The aproximations
which we have done here, sacrificing covariance in a controled way,
lead us to this  calculational scheme where everything is defined. 
Covariant formalisms like those used in \cite{25,26} generate some off shell
dependence in the hadronic tensor, which are
accounted for  in terms of new structure functions for
which there is no empirical information, so several different assumptions
are made in  \cite{25,26} which produce moderate changes in the
EMC  results.

Our relativistic corrections thus stem from the consideration of special
relativity in the positive energy sector, although, as we discussed,
it accounts in an approximate  way for the contributions involving
negative energy states, which are small anyway.

We will show results both with the relativistic and non-relativistic
formalism. The relativistic corrections are found to be relevant
in the region of $x > 1$, particularly at large values of $x$.

It is easy to see qualitatively why eq.(8) or
eq.(19) lead to a non-vanishing structure function for 
$x > 1$. The nucleon structure function appears with argument $x_N$
in this equation and in the Bjorken limit one has

\begin{equation}
x_N = \frac{x_N}{x} x = \frac{M}{\omega - p_{z}} x
\end{equation}

For certain combinations of $\omega$ and $p_{z}$ one can obtain values of
$x_N < 1$ even if $x >1$. Since both $\omega$ and $p_{z}$ appear in
eq.(20)
it is very important to take into account the correlations between
$\omega$ and $p$ provided by the spectral function $S_h (\omega,p)$, and
one sees that approximations which neglect these correlations
are bound to provide unrealistic results.

\section{The nucleon spectral function in nuclear matter and finite nuclei}

We have used three different approaches to evaluate the spectral 
function. The first one is a semiphenomenological one relating the
spectral function of nuclear matter to the experimental cross section
for NN scattering.
The second approach is microscopic in the sense that the spectral
function for nuclear matter is derived from a many-body calculation
employing a realistic One-Boson-Exchange model for the NN interaction.
In the third model we avoid the local density approximation and
evaluate the spectral function ${\cal S}_{h}$ directly for finite
nuclei.
We briefly describe these models below.

\subsection{Semiphenomenological approach}

This model is described in detail in ref.\cite{27}. It evaluates
$Im \Sigma (\omega, p)$ from a second order diagram and uses the fact
that ladder diagrams evaluated from the $NN$ potential lead to the
$NN$ $t$ matrix. Pauli blocking corrections are taken into account in the 
explicit diagram evaluated and $|t|^2$, which appears in the evaluation of
$Im \Sigma$, is written in terms of the experimental $NN$ cross section.
Polarization effects from the $RPA$ iteration   of $ph$ and $\Delta
h$ excitations are also taken into account. The real part of the
selfenergy is obtained via a dispersion relation and the Fock term from
pion exchange is also included. Hartree terms, which require the explicit
knowledge of a potential, are missing in the approach but these are
terms independent of energy and momentum. The nucleon properties evaluated in
\cite{27} as a function of $\omega - \mu$ compare favourably with those
of more microscopic evaluations \cite{28,29}. In order to complete the
model and obtain absolute values for $Re \Sigma$, another phenomenological
piece is added here. This Hartree contribution is assumed to be
proportional to $\rho$ and its value is adjusted, in order to
fit the empirical value of the binding energy per nucleon in each particular
nucleus. For this purpose we recall the sum rule for the binding energy
per nucleon \cite{30}

\begin{equation}
|\varepsilon_A| = - \frac{1}{2} \left( < E - M > + \frac{A - 1}{A - 2}
< T > \right)
\end{equation}

\noindent
and we evaluate $< T >$ and $< E >$ as

$$
< T > = \frac{4}{A} \int d^3 r \int \frac{d^3 p}{(2 \pi)^3}
( E (\vec{p}) - M) \int_{- \infty}^{\mu} S_h (\omega, p) d \omega
$$

\begin{equation}
< E > = \frac{4 }{A} \int d^3 r \int \frac{d^3 p}{(2 \pi)^3} 
\int_{ - \infty}^{\mu} S_h (\omega, p) \omega \, d\omega
\end{equation}

By means of this, one takes also into account empirically contributions
from terms like in fig. 2e, which are in principle small, and even if
they are of $\rho^2$ type would not differ
appreciably from a $\rho \, \rho_{eff}$ form, with $\rho_{eff}$ an
effective average nuclear density.

The evaluation of $Im \Sigma  (\omega,p)$ was done nonrelativistically in
\cite{27}. For consistency with the relativistic formalism used here we 
should have kept the factors $\frac{M}{E}$ in the nucleon propagators
evaluating  $Im \Sigma (\omega,p)$ in ref. \cite{27}.
However, the range of momenta in the loop integrals in $Im \Sigma
(\omega,p)$ is quite limited and  they would modify the values
of $Im \Sigma (\omega,p)$ by less than 10$\%$.
By means of the explicit calculations carried out here we have observed
that an increase of 10 $\%$ in $Im \Sigma (\omega,k)$ leads to 
increases of $F_{2A} (x)$ of the order 2 $\%$ at $0 < x <0.6$ and always
below the 10 $\%$ level for large $x$, hence we have continued to use the same
$Im \Sigma (\omega,p)$ as obtained in ref.\cite{27}.

\subsection{Microscopic approach in nuclear matter}

The spectral function of nuclear matter which has been used in this
second approach has been evaluated using the techniques described in
ref.\cite{31}. The starting point of this many-body calculation is a
Brueckner-Hartree-Fock (BHF) calculation of nuclear matter considering the
realistic One-Boson-Exchange (OBE) potential $B$ as defined in \cite{32} for
the NN interaction. The $G$-matrix resulting from this BHF calculation
as well as the BHF single-particle spectrum $\epsilon_{BHF}(p)$ are used
to define the nucleon self-energy including all terms up to second order
in $G$. The single-particle Green's function $g(p,\omega)$
is derived from a solution of the Dyson equation

\begin{equation}
g(p,\omega ) = g^{(BHF)} (p,\omega ) + g^{(BHF)} (p,\omega ) \left[
\Sigma^{(2)}(p,\omega ) \right] g(p,
\omega)\, 
\end{equation}

Here $\Sigma^{(2)}$ is the contributions to the self-energy in second
order. The single-particle Green's function in the BHF
approximation is given by

\begin{equation}
g^{(BHF)} (p,\omega ) = \frac{\Theta (k_{F}-p)}{\omega -\epsilon_{p}
-i\eta } + \frac{\Theta (p -k_{F})}{\omega -\epsilon_{p} + i\eta }\, ,
\end{equation}

\noindent
where $k_{F}$ denotes the Fermi momentum of nuclear matter at the
density under consideration. The term with $\Sigma^{(2)}$ in eq. (23)
contains a contribution with intermediate particle-particle states,
which is taken into account already in the BHF aproximation.
This doublecounting is removed as described in
\cite{31}.
The spectral function $S_{h}(\omega,p)$ can then be calculated from the 
imaginary part of the single-particle Green's function by

\begin{equation}
S_{h}(\omega,p) = \frac{1}{\pi}  Im \, 
g(p, \omega ) \, 
\end{equation}

This calculation yields a Fermi energy $\mu$ depending on the density
of nuclear matter. The energy variable $\omega$ is defined with respect
to this Fermi energy. In the local density approximation for the
spectral function discussed above, the empirical Fermi energy of the
finite nucleus has been chosen to be the reference point for the energy
variable $\omega$.

\subsection{ Microscopic approach for finite nuclei}

The spectral function can be calculated directly for finite nuclei
using the procedure described and applied to $^{16}$O in ref.\cite{33}.
For nuclei with spherical symmetry the self-energy is evaluated in a
partial wave basis, $\Sigma_{lj}(p,p')$, assuming that orbital angular
momentum $l$ and total angular momentum $j$ are conserved quantum
numbers. As discussed above, the total self-energy is decomposed in a
BHF part and terms of second order in the Brueckner G-matrix. The
corresponding single-particle Green's function can be evaluated by
solving a Dyson equation of the form

\begin{eqnarray}
g_{l,j}(p,p';\omega ) =  g^{(BHF)}_{l,j} (p,p';\omega ) + & \int dk_{1}
\int dk_{2}\ g^{(BHF)}_{l,j} (p,k_{1};\omega ) \nonumber\\ & \ \times \left[
\Sigma^{(2)}_{l,j}(k_{1},k_{2};\omega ) \right] g_{l,j}(k_{2},p';
\omega)\, 
\end{eqnarray}

The spectral function for the various partial waves is then obtained
from the imaginary part of the Green's function $g_{l,j}(p,p;\omega)$
applying eq.(25). A problem of this partial wave expansion
for the momentum distribution is related to the fact that
non-negligible contributions are obtained at large momenta and energies
in high partial waves. Therefore we prefer to apply an approach which
has been introduced and discussed in ref.\cite{31}. In this
approximation one splits the spectral function, for nuclear matter as
well as for finite nuclei, into a quasiparticle contribution describing
the contribution to the spectral function around the quasiparticle pole
and a background contribution which contains the information about the
spectral function at energies away from the respective quasiparticle
pole. For finite nuclei a quasiparticle pole contribution is only
observed for those partial waves, which are occupied in the HF or
independent particle model. Therefore the sum on partial waves in

\begin{equation}
S^{QP}(\omega,p) = \sum_{l,j} 2(2j+1) n_{l,j} \delta(\omega -
\epsilon_{l,j}^{QP}) \vert \Phi_{l,j}(p) \vert^2
\end{equation}

\noindent
is restricted to $l$=0 and 1 in our example of $^{16}O$. In this
equation $\epsilon_{l,j}^{QP}$ stands for the energy of the
quasiparticle pole, $\Phi_{l,j}(p)$ for the corresponding
single-particle wave function in momentum space and $n_{l,j}$ for the
occupation probability for this pole. This quasiparticle pole
contribution is supplemented by the background contribution calculated
in a local density approximation from the background contribution in
nuclear matter

\begin{equation}
S_{h,A}(\omega, p) = S^{QP}(\omega,p) + 4 \int d^3r S_{h}^B
(\omega,p;\rho(r))\label{eq:sfin}
\end{equation}

\noindent
where $S_{h}^B$ stands for the background contribution of the
spectral function calculated in nuclear matter at the local density
$\rho(r)$. Care is taken that the whole spectral function is normalized
such that

\begin{equation}
\int \frac{d^3 p}{(2 \pi)^3} \int_{-\infty}^\mu d\omega
S_{h,A} (\omega,p) = A
\end{equation}

with $A$ the total number of nucleons.

\subsection{Approximations to be avoided}

If no reliable model for the spectral function for nucleons in nuclear
matter is available, one may be tempted to use certain approximations.
One of such approximations, which has frequently been used
\cite{11,12,13} is to ignore the special correlations between momentum
and energies of nucleons provided by the spectral function and simply
use the energy-integrated spectral function, which is the momentum
distribution. We are going to discuss three different approximations
and try to explore their reliability by comparing with the
results obtained in the more sophisticated models for the spectral
function discussed above.  Although discussions around different 
approximations to the nuclear wave functions, and other different 
approximations, have been common in the past \cite{34,2,4,7},
the comparison of different approximations to the results obtained using
spectral functions has not been exploited, particularly in the region
of $x > 1$ which we study here.

\vspace{0.5cm}

\noindent
{\it a) The uncorrelated Fermi sea distribution.}

\noindent
This is the simplest 
approximation, which is usually very accurate, except of course in processes
which test the momentum distribution at momenta which are large compared to
the Fermi  momentum, as is
the case in the present problem.
In this approximation the spectral function within the local density
approximation is assumed to have the form

\begin{equation}
S^{UFS}_{h} (\omega, p;\rho )  = n_{FS} (\vec{p}) \delta (\omega
 - E (\vec{p}) - \Sigma)
\label{eq:apufs}
\end{equation}

\noindent
with an occupation probability of

\begin{equation}
n_{FS} (\vec{p}) = \cases{1 & if $|\vec{p}| < p_F (r)$\cr
0 & if $|\vec{p}| > p_F (r)$} \label{eq:mm1}
\end{equation}

\noindent
with a local Fermi momentum $p_{F}(r)$ which is related to the local
density $\rho(r)$ by

\begin{equation}
p_F (r) \equiv (\frac{3 \pi^2 \rho (r)}{2})^{1/3}
\end{equation}

\noindent
and an expression for the local single-particle potential

\begin{equation}
\Sigma \equiv \Sigma (r) = V_{TF} (r) + D \rho (r)
\end{equation}

\noindent
where $V_{TF} (r)$ is the Thomas Fermi potential, $- p_{F}^2 (r)/ 2M$ and
$D$ a
phenomenological constant fitted to reproduce the binding energy 
per nucleon in the nucleus, as done in section 3.1.

\vspace{0.5cm}

{\it b) Use of the momentum distribution of the correlated Fermi sea.}

Since large momentum components are needed to generate $F_{2A} (x) $ at
$x > 1$, one is tempted to use the realistic momentum distribution of the
nucleus as a way to improve on this approximation.
This means that we assume an expression for the spectral function
$S^{MD}_{h}$ like in eq.(\ref{eq:apufs}) but replace the momentum
distribution of the free Fermi gas $n_{FS}$ by the momentum
distribution of the interacting Fermi gas  

\begin{equation}
n_{I} (\vec{p}) = \int_{- \infty}^\mu S_h (\omega, p) d \omega
\label{eq:mm2}
\end{equation}

\noindent
where we have used the spectral function of nuclear matter discussed
above to calculate $n_{I}$. Note, however, that the energy-momentum
relation is still determined by the $\delta$-function in
(30), with $\Sigma$ defined with the same prescription
as in the subsection above, eq. (33).

\vspace{0.5cm}

{\it c) Use of the correlated momentum distribution and the corresponding mean
value for the energy.}

Finally we want to consider an approximation in which we assume again a
definite relation between momentum and  energy of a nucleon in
the hole spectral function

\begin{equation}
S_{h}^{MED} (\omega,p;\rho ) = n_{I} (\vec{p}\,)  \delta (\omega - 
< \omega (\vec{p})>)
\end{equation}

\noindent
but determine the momentum distribution $n_{I}$ (see eq.(34))
as well as the mean value of the energy for a given momentum

\begin{equation}
< \omega (\vec{p}\,)> = \frac{\int_{- \infty}^{\mu} S_h (\omega,p) 
\omega d\omega}{ \int_{- \infty}^\mu S_h
(\omega, p) d\omega } \label{eq:wmean}
\end{equation}

\noindent
from the complete spectral function for nuclear matter discussed above.

\section{Results and discussion}

In a first step we want to compare the two approaches to determine the
spectral function for nuclear matter, which we describe in sections 3.1
and 3.2 and which we are going to employ for the calculation of
the structure function. For that purpose we present in Fig. 4 the
momentum distribution $n_{I}(p)$ (see eq.(34)) calculated at
the empirical saturation density, $\rho = \rho_0$, of nuclear matter.

The momentum distributions obtained by these two very different methods
are very similar. At small values of  $p$
the microscopic approach of \cite{31} provides an occupation number
of the order of 3 $\%$ bigger than  the semiphenomenological one \cite{27}.
For momenta above the Fermi momentum the distributions are also similar
although for momenta around two times the Fermi momentum ($\sim 550 \, MeV/c$)
the differences become more appreciable. The semiphenomenological approach
provides a little less strength below the Fermi momentum, 
which is then redistributed to larger momenta where 
$n_{I} (\vec{p}\,)$ is larger than in the microscopic approach. The precise
values of $n_{I} (\vec{p})$ calculated in a microscopic many-body
theory, depend on the model of the NN interaction
which is considered  and the method which is used to determine the
effects of correlations.
For instance in the self-consistent Green's function approach of ref.
\cite{29}, using the Reid soft-core potential, the
occupation number for momenta below $k_{F}$ is around 0.85, smaller
than in both the approaches considered here. This demonstrates that
the semiphenomenological approach provides a result which is in agreement
with the microscopic calculations within the uncertainties of the
microscopic approach caused by the approximation in the many-body
theory as well as NN interaction.
The differences found in 
$n_{I} (\vec{p})$ in the two approaches discussed have little 
repercussion in the values of $F_{2A} (x)$, which come very close
to each other in the two approaches, as we shall see below.

As a second quantity characterizing the bulk properties of the spectral
functions calculated by these two methods, we show in Fig. 5 the mean
value for the energy as a function of $\vec{p}$ calculated according to
eq.(36) at the nuclear density $\rho_{0}$. 
 Fig. 5 shows in a qualitative way that there
is an important correlation between the momenta and the mean value of
the energy for the bound nucleons. The absolute value of this mean energy
$|< \omega (\vec{p} ) - M>|$ decreases as a function of momentum with
increasing momenta for momenta below the Fermi momentum. This momentum
dependence is mainly due to the momentum of the quasiparticle peak,
which is approaching the Fermi energy for $p \to p_{F}$. There is no
quasiparticle contribution to the hole-spectral function $S_{h}$ for
momenta larger than the Fermi momentum $p_{F}$. Therefore at these
momenta, the mean value is determined only from the background
contribution. The coupling to 2 hole-1 particle and more complicated
configurations with total momenta $p$, described  by these background
terms yields a mean value of $<\omega - M> $, which decreases with
increasing momentum.
From this figure it is evident, however, that the energy-momentum
relation obtained from a realistic spectral function is quite different
from the simple relation used in eqs. (30), (33), which provides a dispersion
relation which is always an increasing function with increasing
momentum.

The values obtained for $|< \omega (\vec{p})>|$  in the two approaches
discussed in sections 3.1, 3.2
are very similar, with differences of the order of 10 $\%$ at
most. This is another indication that the basic features of the
spectral function are not very sensitive to the method used in the
evaluation and that also the semiphenomenological approach yields quite
a reliable result.

In Fig. 6 we show the results for $F_{2A} (x)$ calculated with the
three different spectral functions introduced in sections 3.1 to 3.3,
 for the case of 
$^{16}O$. The density distributions $\rho(r)$ for $^{16}O$  and the
other nuclei, which are required to apply the local density
approximation, are taken from
refs. \cite{35,36}. Since the microscopic nuclear matter and finite nuclei
approaches are nonrelativistic, we have also taken the nonrelativistic 
version of the semiphenomenological approach, for comparison, 
omitting all the $\frac{M}{E (p)}$ factors in eqs. (17),
(19). The
experimental values for $F_{2N} (x_N)$ are taken from ref. \cite{37}.
The results obtained with the two spectral functions of nuclear matter
(solid line and dashed line) are rather
similar. At $x \simeq 1$ the microscopic spectral function provides
results about 20 $\%$ higher than the semiphenomenological one. At 
values of $x \simeq 1.22$ the two approaches coincide and for $x \simeq
1.5$, where the structure function has decreased three orders
of magnitude with respect to the value at $x = 1$,
the semiphenomenological approach
provides values of $F_{2A}$ about 40 $\%$ larger than the  microscopic
one. This reflects the fact that the former model provides a larger
probability for the momentum distribution at high momenta 
than the latter one, as seen in Fig. 4. 

The results for the structure function obtained with the spectral
function of eq.(28) evaluated directly for the finite
nucleus are represented by the dot-dashed line in Fig. 6.
They should be compared with those displayed by the solid line since
the background contribution to eq.(\ref{eq:sfin}) is obtained from the
same nuclear matter result. These two results can hardly be separated
on the logarithmic scale of the figure. We observe that at $x \simeq 1$ the
results with the spectral function of the finite nucleus are about 8 $\%$
bigger than with the nuclear matter approach. The differences 
become smaller as $x$ increases and for values of $x \simeq 1.5$ the
two approaches give the same results. This latter fact is telling us 
that at large values of $x$ one is getting practically all contributions
from the background part of the spectral function and none from the 
quasiparticle part. The comparison of these
two curves also tells us that the use of the nuclear matter spectral
functions, together with the local density approximation, is
an excellent tool to evaluate $F_{2A} (x)$. If one compares the results
at values of $x$ studied in the $EMC$ effect, $0 < x < 0.6$, the differences
among the three calculations are of the order of 3 $\%$.

In Fig. 7 we show results obtained with the semiphenomenological approach
using the relativistic and nonrelativistic formalisms. The trend of the 
results is similar, however, the relativistic corrections induce a 
reduction of 25 $\%$ around $x = 1$ and roughly reduce the structure
function $F_{2A}$ to one half of the
non relativistic results at $x \simeq 1.5$. The relativistic 
effects are significant in the sense that they are bigger than 
the differences found between various  nonrelativistic approaches, which
reflect the uncertainties in the treatment of correlations.

Results obtained for $F_{2A} (x)$ using  the 
different approximations discussed in section 3.4 are displayed in
Fig. 8. The first one, which originates from the assumption of an 
uncorrelated Fermi sea, eq. (30), is represented by the
dot-dashed line. We can see
that at $x \simeq 1$ it already provides a structure function of
around a factor two smaller than the one obtained with the proper spectral 
function (short dashed-line). However, as one moves to higher $x$,
the discrepancies get bigger
and at values of $x \simeq 1.2$ the uncorrelated Fermi sea gives
already values for the
structure function
 which are about two orders of magnitude smaller than the
correct ones. It is clear that one is exploring the region of large
momenta, above the Fermi momentum, which are not accounted for by 
the uncorrelated Fermi sea.

Another approximation corresponds to using the realistic momentum 
distribution $n_{I} (\vec{p})$ of eq. (34) and associating an
energy to
each $\vec{p}$ given by its kinetic energy plus a potential, eq.
(33). The results (solid line) are outrageously wrong. This
demonstrates that the naive use of a momentum distribution, although
calculated in a realistic way, may lead to results which are worse than
those obtained for an uncorrelated system, if one does not treat the
energy-momentum correlation properly. As we have discussed already in
Fig. 5, the mean value of the energy $< \omega (\vec{p}\,)>$ decreases
above the Fermi momentum with increasing momenta. On the
other hand, the energy associated to $\vec{p}$ in eq. (30)
grows like
the kinetic energy as $|\vec{p}|$ increases. The discrepancies with the
correct results are about a factor three at $x \simeq 1$ and three
orders of magnitude at $x \simeq 1.5$, providing a gross overestimate
of the results for $F_{2A} (x)$. The same gross overestimate found
here for this approximation was also found in ref. \cite{14} in 
connection with the mesonic Lambda decay in nuclei.

In view of the deficiencies of the previous approximations and the 
reasons for it, one might think that the results should be improved by
replacing the kinetic plus potential
energy, eq. (30), by the mean value of $< \omega 
(\vec{p}\,)>$ calculated from the spectral function (see
eq.(36). This is indeed the case (see curve with long dashes
in Fig. 8), although the discrepancies with the exact results are still
large 
enough to discourage this approximation too. We can see in Fig. 8
that at values of $x \simeq 1$ (and also in the EMC region below)
the approximation turns out to be quite good. However, for values of
$x \simeq 1.3$ and above the discrepancies with the
correct results are already as big as one order of magnitude or more.

The results discussed here stress the importance of using the spectral
function to evaluate $F_{2A} (x)$ since all the information contained in it, 
correlating energies and momenta, is very important, particularly at large $x$. We 
showed that some schemes which use only a partial information from the
spectral function lead to rather inaccurate results and
should thus be avoided.

Finally in Fig. 9 we show results obtained for different nuclei. 
They are calculated at $Q^2 = 5 \; \,  GeV^2$. We can see
that $F_{2A} (x) /A$ is very similar for the different nuclei. We have
taken nuclei with $N =Z$ or close by, to be able to use a unique Fermi
sea for protons and neutrons as done in symmetric nuclear matter. For 
heavier nuclei with $N \neq Z$ the results obtained here could be easily
extended  by dealing with two different Fermi seas, but this would require
the extension of the work of ref. \cite{27} to non symmetric nuclear
matter. We do not expect, however, any special effects apart from those
exposed here.

When evaluating absolute values of nuclear structure function, not
ratios to the nucleon or the deuteron, it is very important to
take into account the $Q^2$ dependence of the structure function. This
is particularly true for values of $Q^2 \simeq 1 - 10 \,\; GeV^2$,
but even at $Q^2 \simeq 100 \,\; GeV^2$ and above, where there is
approximate Bjorken scaling, the $Q^2$ dependence is weak but one
still has to consider it if one wants to make accurate predictions.

For the $Q^2$ dependence of the nucleon structure function 
we have taken the parametrizations given in ref. \cite{37A}.

It is interesting to compare our results with the scarce  experimental
data available. Our  results refer exclusively to the deep inelastic
contribution to electron nucleus scattering. The quasielastic 
contribution (where only one nucleon is knocked out in the first
step $e N \rightarrow e' N$) is not taken into account in our formalism.
At low values of $Q^2$ and $x \geq 1$, the quasielastic contribution
is dominant \cite{15C} and one has to go to values of $Q^2 > 20 \, \; GeV^2$
to have a dominance of the deep inelastic contribution \cite{15A}. For
this reason we compare our results with measurements done at $Q^2 = 61,
85$ and $150 \, \; GeV^2$ in ref. \cite{37B}, which improve the preliminary
results reported in ref. \cite{38} where much larger values were obtained.

The results can be seen in fig. 10. The three theoretical curves
correspond to each one of the values of $Q^2$
and the results decrease as a function of increasing $Q^2$.
 The agreement with the
data is qualitative. The slope as a function of $x$ seems well reproduce
but the theoretical results are in average 40 $\%$ higher than experiment
up to $x = 1.05$. At $x = 1.15$ and $1.3$ there are only upper bounds 
which are compatible with our predictions.

Experimental results at $Q^2 < 5 \, \; GeV^2$ have however, a large contamination of
quasielastic contribution \cite{15A,15B,15C}. This is  reflected
by the large dispersion of the  results as a function of $Q^2$ \cite{39}
and in the approximate  $y$
scaling of these results,  which is characteristic
of the quasielastic collisions. Nevertheless, we have also
evaluated the deep inelastic contribution correspondieng to the 
results in \cite{39} with largest values of $Q^2$. We evaluate the
structure function corresponding to the lowest curve in fig. 1 of 
ref. \cite{39}. This corresponds to different values 
of $Q^2$ for each value of 
$x$ since the data correspond to electron scattering with fixed
initial electron energy ($E_e = 3.595 \,\; GeV$), fixed scattering
angle, $\theta = 39^0$ and variable final electron energy. We show the
results in fig. 11. The values of $Q^2$ increase with increasing 
$x$.
At $x = 1 \, , Q^2 = 3.11 \,\; GeV^2$ and at $x = 1.25 \, ,
Q^2 = 3.42$. We can see that our results  lie below the experimental
data, particularly at large values of $x$. However, one can observe
a tendency to be in agreement with the data at values of
$x < 0.8$ if one extrapolates the data smoothly. In fact
in fig. 1 of ref. \cite{39} we see a confluence of the data 
for different values of $Q^2$ in the region of $x = 0.4 - 0.6$ with values
which agree with our results of fig. 11.
  This would be in agreement with the conclusions reached in 
\cite{39} where the large dispersion of the results as a function
of $Q^2$ for large values of $x$ indicates the dominance of the 
quasielastic contribution, while the tendency to stabilize the 
results at values of $x < 0.8$ indicate that this region of $x$ is
dominated by the deep inelastic contribution. In such case our results
should be comparable to the data and this is
indeed the case.

\section{Conclusions}

We have evaluated the nuclear structure function $F_{2A} (x)$ at values
of $x$ bigger than unity, especially in the range $1 < x < 1.5$ where
the values obtained are well within measurable range. For this
purpose we have used sophisticated nuclear spectral functions
which account for nuclear correlations and relativistic
effects. The structure function 
decreases three orders of magnitude from $x = 1$ to $x = 1.5$.

The strength of $F_{2A} (x)$  in that range of $x$ is tied to the
components of the nuclear wave function with nucleons of large momenta.
These components are due to the two-nucleon correlations originating
from realistic NN interactions. The momentum distribution of nucleons
in the nuclear many-body system, however, is
strongly correlated with the energy distribution of these nucleons. These
are dynamical effects which go beyond the shell model picture of the
nucleus and which are taken into account in terms of the nucleon spectral
functions. The results for $F_{2A} (x)$ are very sensitive to the 
correlations between $\omega$ and $p$, to the extreme that usual approximations
made in calculations of the $EMC$ effect fail badly in the region of
$x > 1$. In particular, at $x \simeq 1.5$ the quasiparticle bound states (the
occupied states of the shell model), which are only partly occupied 
in an interacting nucleus, give a negligible contribution to $F_{2A} (x)$
and all the strength comes from the background part of the spectral
function.

We have discussed in detail the results obtained with several
approximations which use only rough spectral functions or partial information
from realistic ones, and which 
are often used. We showed that in this region of
$x$ none of them can be taken as a substitute of the  calculation
using the whole information of the spectral function.

In order to quantify the intrinsic theoretical uncertainties of the
results we used two different models for the spectral function 
evaluated in infinite nuclear matter and $F_{2A} (x)$ for nuclei
was calculated using the local density approximation. A 
version of the spectral function for finite nuclei was used also for
$^{16}O$. The
differences between the models were small, of the order of 10-30 $\%$
depending on the region of $x$. We also found that the use of the
local density approximation was an excellent tool, providing results
very close to those obtained by direct evaluation for the finite nucleus.

Relativistic effects were checked and found to be important. They
reduce the results for $F_{2A} (x)$ obtained with the nonrelativistic 
approximation by amounts ranging from 25 $\%$ at $x \simeq 1$ to 
nearly a factor two at $x \simeq 1.5$.

On the other hand we have evaluated $F_{2A} (x)$ for different nuclei and
find that $F_{2A} (x) /A$ becomes very similar
for $N = Z$  nuclei from $^{40}Ca$ on.

The experimental results at $x > 1$ are scarce, particularly at large
values of $Q^2$. We compared our results with available data
at $Q^2 = 61, 85, 150 \,\; GeV^2$ and found our results about 40$\%$
higher than experiment, although the fall down with $x$ was well 
reproduced. At values of $Q^2$ significantly smaller, $Q^2  < 4 \,\; GeV^2$,
we found that our results for $x > 1$ where much smaller than the 
experimental data, which was in agreement with theoretical and
experimental findings that this region is dominated by quasielastic 
scattering. At lower values of $x$, around $x = 0.4 - 0.6 $ 
 our results  matched the experimental data, in 
agreement with the theoretical and experimental findings that this
region is dominated by the deep inelastic contribution.

The present investigation and the importance of the nucleon spectral function
for the precise determination of $F_{2A} (x)$ is telling us that measurements 
of this quantity for different values of $x$ and a wide range of nuclei
would provide important information on the components
of the nuclear wave function at large momenta and energies and
the strong correlations between momenta and energy. This
information would be very important as a test of the many body theories
which are employed for the determination of the spectral function
and would unveil interesting details on nuclear correlations, 
which complement our knowledge of nuclear structure beyond the
basic information contained in the shell model wave functions.

\vspace{3cm}

Acknowledgments:

We would like to thank discussions with D.B. Day, J. S. McCarthy and
R. Minehart concerning their experiment \cite{39} and with G.I. 
Smirnov concerning the experiment of ref. \cite{37B}. Discussions
with F. Gross and S. Liuti on theoretical issues are equally
appreciated.

One of us, E.O. wants to thank the Humbold Foundation who supported his
stay at T\"{u}bingen. P.F. and E.M. wish to thank the financial
support of the Human Capital and Mobility program of the EU, contract
no. CHRX-CT 93-0323. The work is partly supported by CICYT contract
no. AEN 93-1205.

\noindent
{\bf Figure Captions:}

\vspace{0.4cm}

\noindent
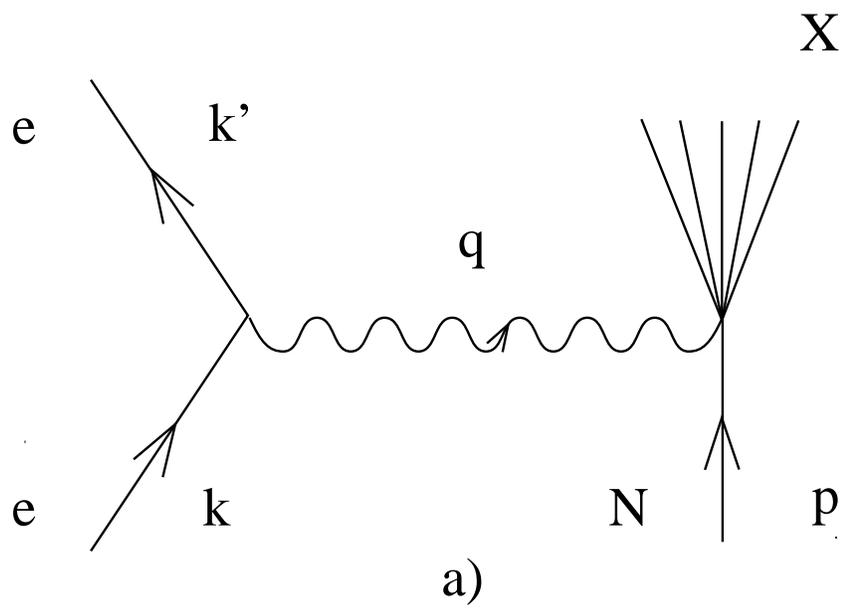
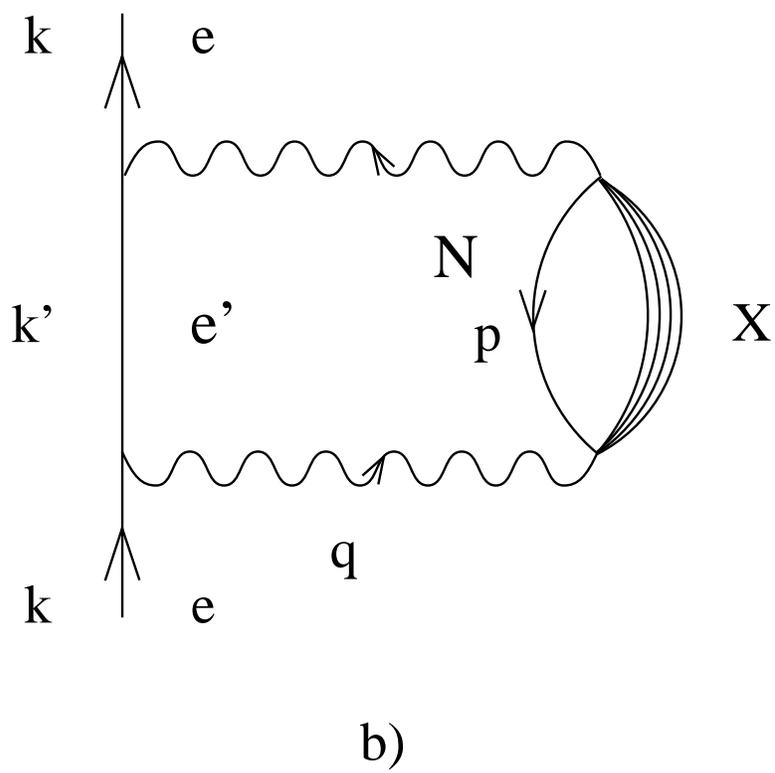
Fig. 1: a) Feynman diagram for deep inelastic electron-nucleon
scattering and, b) electron selfenergy diagram associated to it.

\vspace{0.2cm}

\noindent
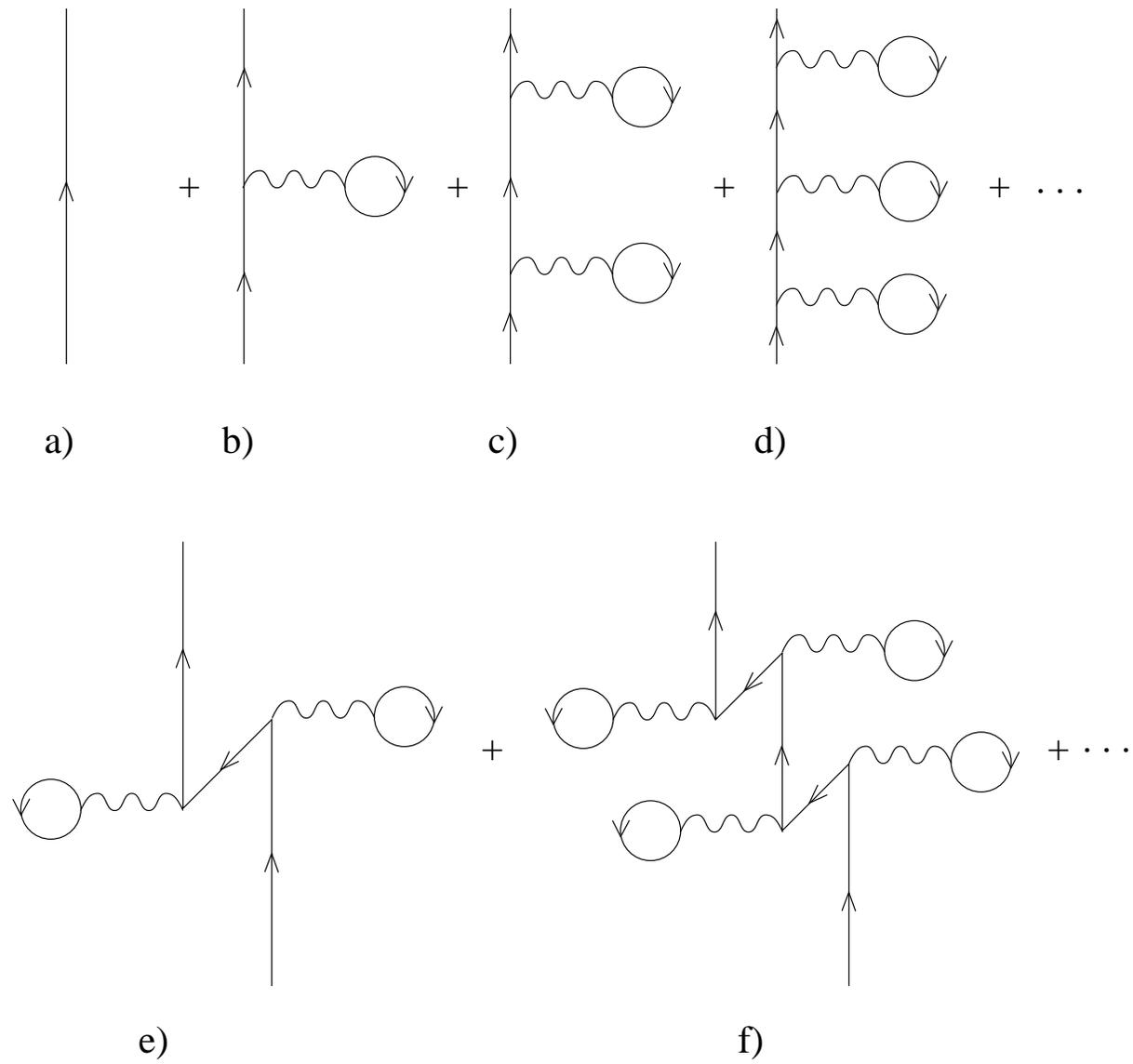
Fig. 2: a,b,c,d) Feynman diagrams of the Dyson series in the 
evaluation of the nucleon propagator including only intermedite 
positive energy states. e) Feynman diagram of the Dyson series  with a
negative energy intermediate state. f) Higher order terms in
the Dyson series originated from the selfenergy term of 
diagrame and its iteration through positive energy intermediate states.

\vspace{0.2cm}

\noindent
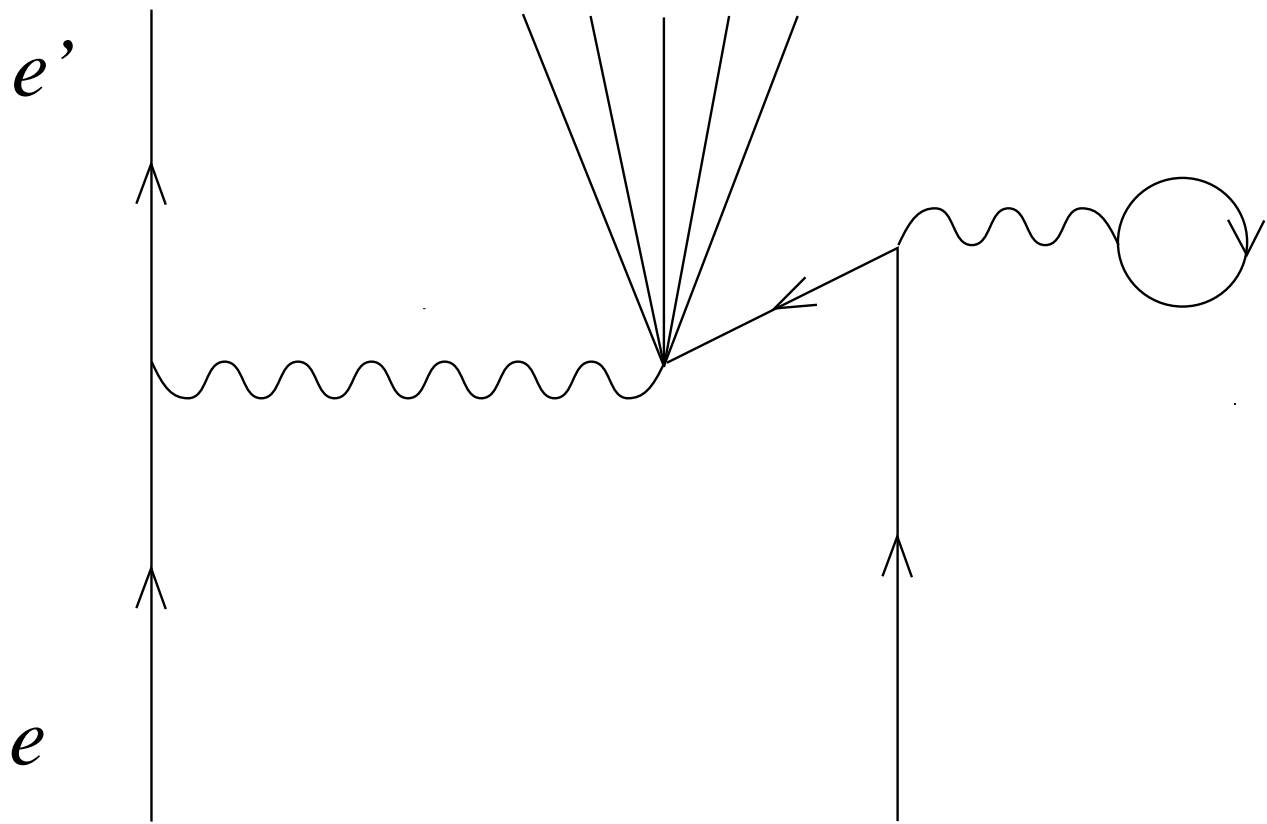
Fig. 3. Feynman diagram for deep inelastic electron scattering with
an interating nucleon, through the coupling of the photon to 
the negative energy components.

\vspace{0.2cm}

\noindent
Fig. 4: Momentum distributions at $\rho = \rho_0$. 
Solid line: microscopic model of 
\cite{31}; dashed line: semiphenomenological model \cite{27}.

\vspace{0.2cm}

\noindent
Fig. 5: Mean value of the energy of nucleons  as a function of $\vec{p}$,
from eq. (36), at $\rho = \rho_0$. Solid line: microscopic model \cite{31};
dashed line: semiphenomenological model \cite{27}.

\vspace{0.2cm}

\noindent
Fig. 6: Results obtained for the structure function of $^{16}O$. 
Solid line: microscopic nuclear matter model \cite{31}; dot-dashed line: 
microscopic finite nuclei model eq.(23); dashed line: nonrelativistic 
semiphenomenological model \cite{27}.

\vspace{0.2cm}

\noindent
Fig. 7: Results obtained for the structure function of $^{16}O$ using
the semiphenomenological model \cite{27}. Solid line: nonrelativistic 
formalism; dashed line: relativistic formalism.

\vspace{0.2cm}

\noindent
Fig. 8: Results obtained for the structure function of $^{16}O$ using 
different approximations. Dot-dashed line: uncorrelated Fermi sea, eq.
(25);  solid 
line: momentum distribution of the correlated Fermi sea, eq. (29);
long dashed line: momentum distribution 
of the correlated Fermi sea
and average energy $< \omega (\vec{p}\,)>$, eq. (30); short dashed line:
spectral function, eq. (13). $Q^2 = 5 \,\;  GeV^2$.

\vspace{0.2cm}
\noindent
Fig. 9: Results obtained for the structure function per nucleon
for different nuclei. Solid line: $^{40}Ca$; short dashed line:
$^{56}Fe$; dot-long dashed line: $^{12}C$; dot-short dashed line: $^{16}O$;
long dashed line: $^6 Li$. All results are obtained using the
relativistic version of the spectral function, eq. (13). $Q^2 = 5 \, \; GeV^2$.

\vspace{0.2cm}

\noindent
Fig. 10: Results for the structure function of $^{12}C$ at $Q^2 = 61, 85 $
and $150 \, \;  GeV^2$ (solid, short-dashed and long dashed lines
respectively). The data are from ref. \cite{37B} crosses for 61 $GeV^2$, 
squares for 
85 $GeV^2$ and triangles for 150 $GeV^2$. The data for the two largest
values of $x$ are upper bounds.

\vspace{0.2cm}

\noindent
Fig. 11:  Results for the deep inelastic structure function of 
$^{56}Fe$ at different values of $Q^2$ around 3 $GeV^2$, see text, 
compared with experimental inclusive data results of \cite{39}.
The experimental data for values of $x >1$ are dominated by the quasielastic 
contribution while for values of $x < 0.8$ the deep inelastic
contribution dominates the reaction.

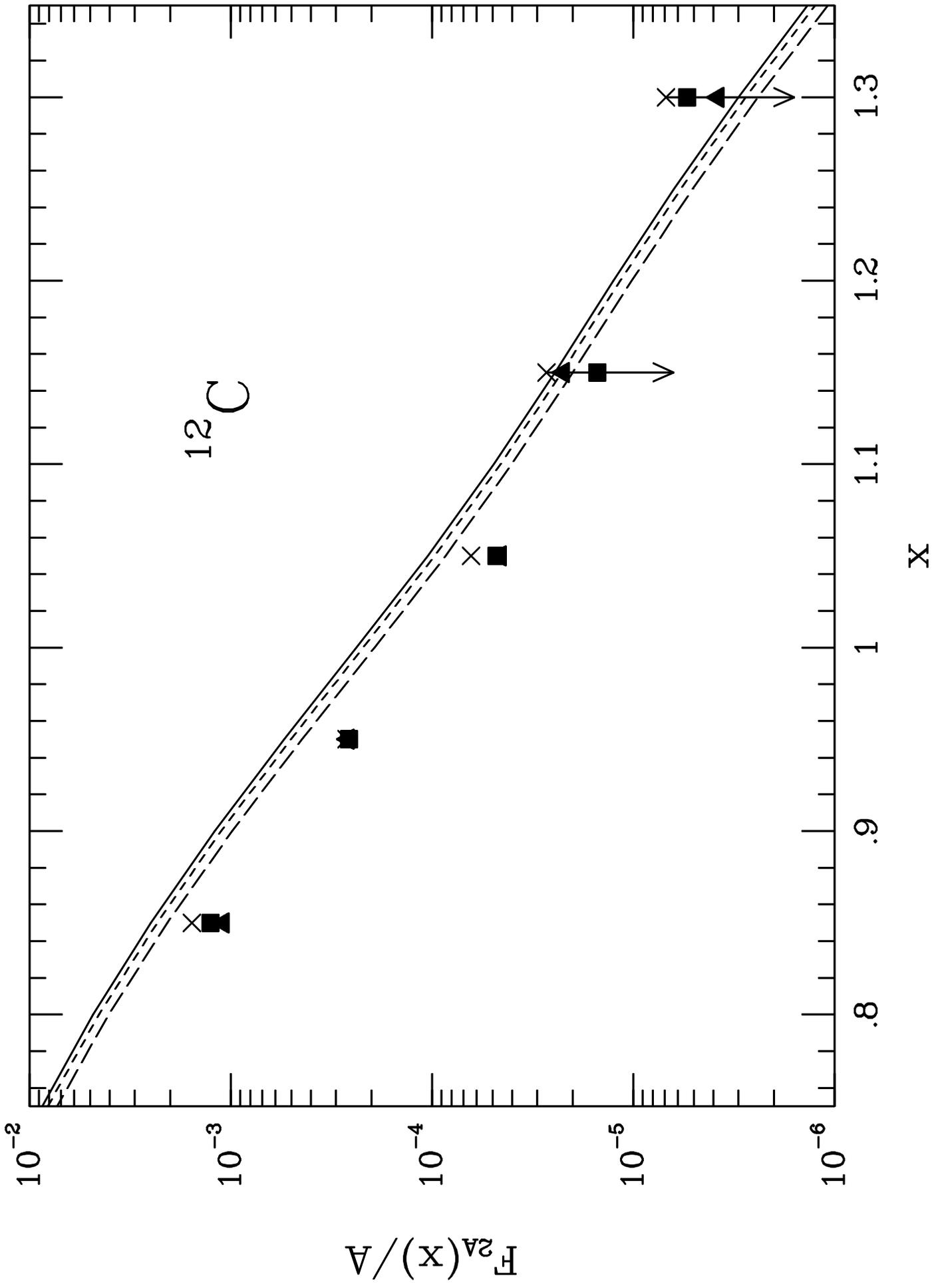
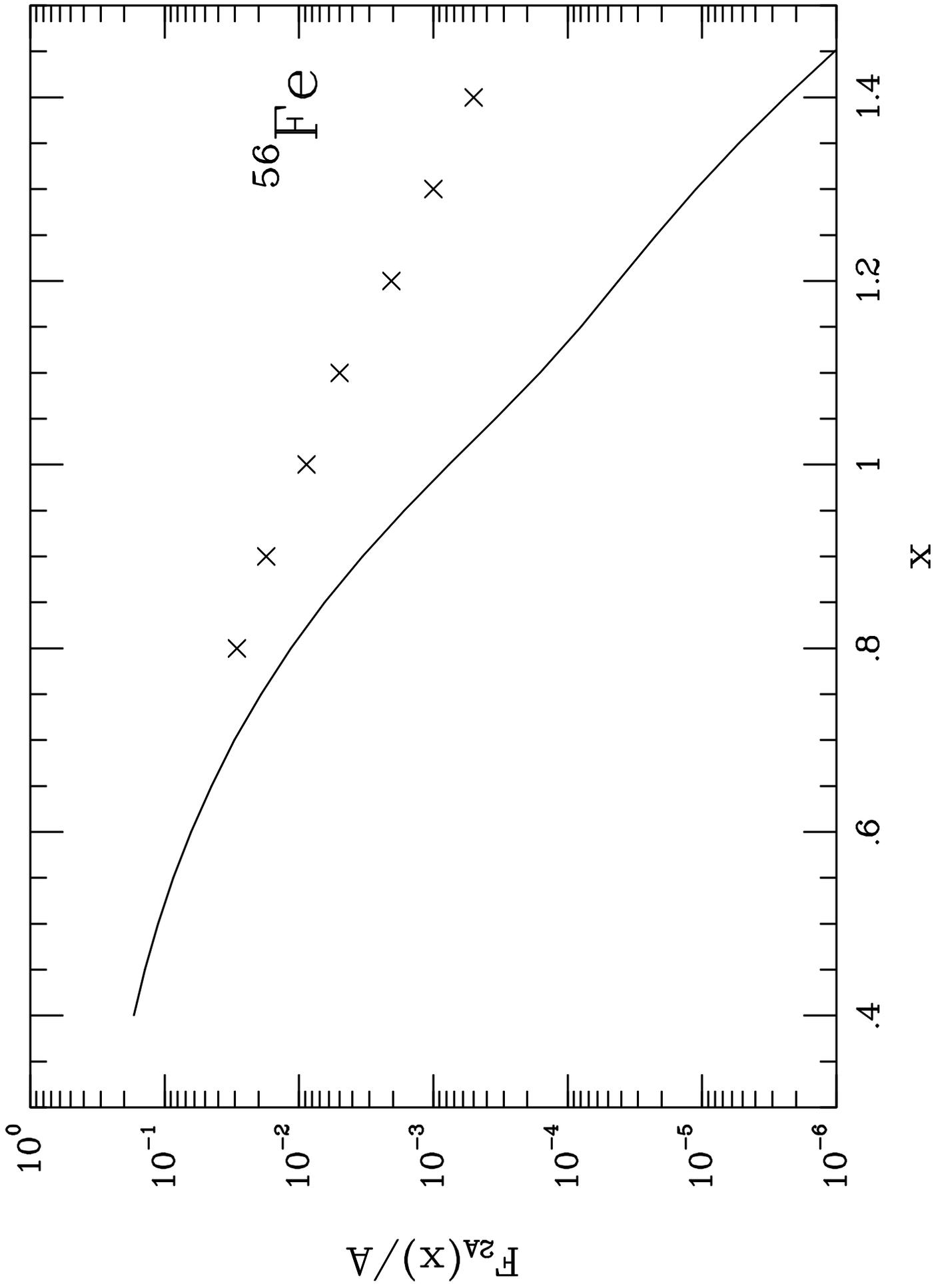
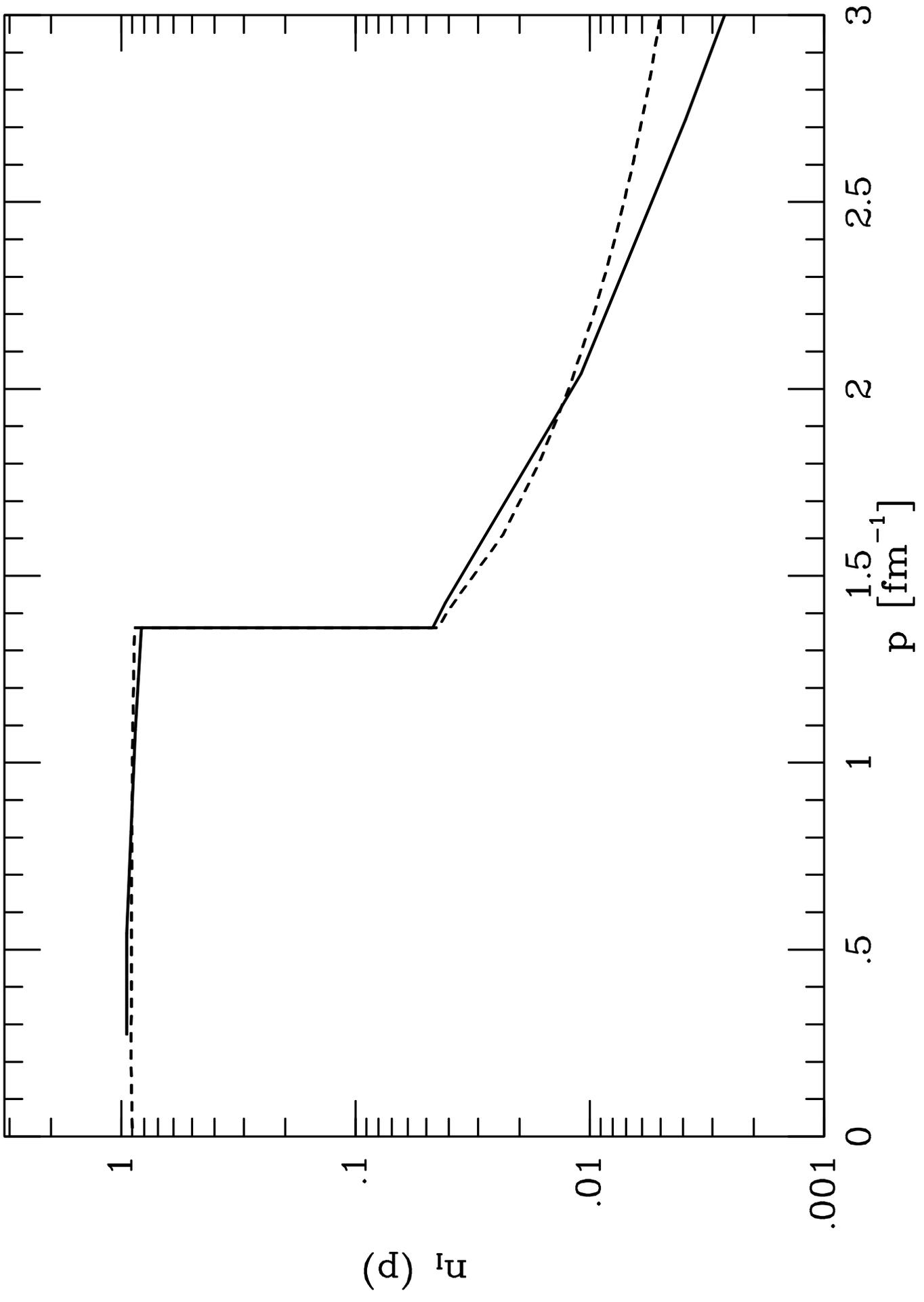
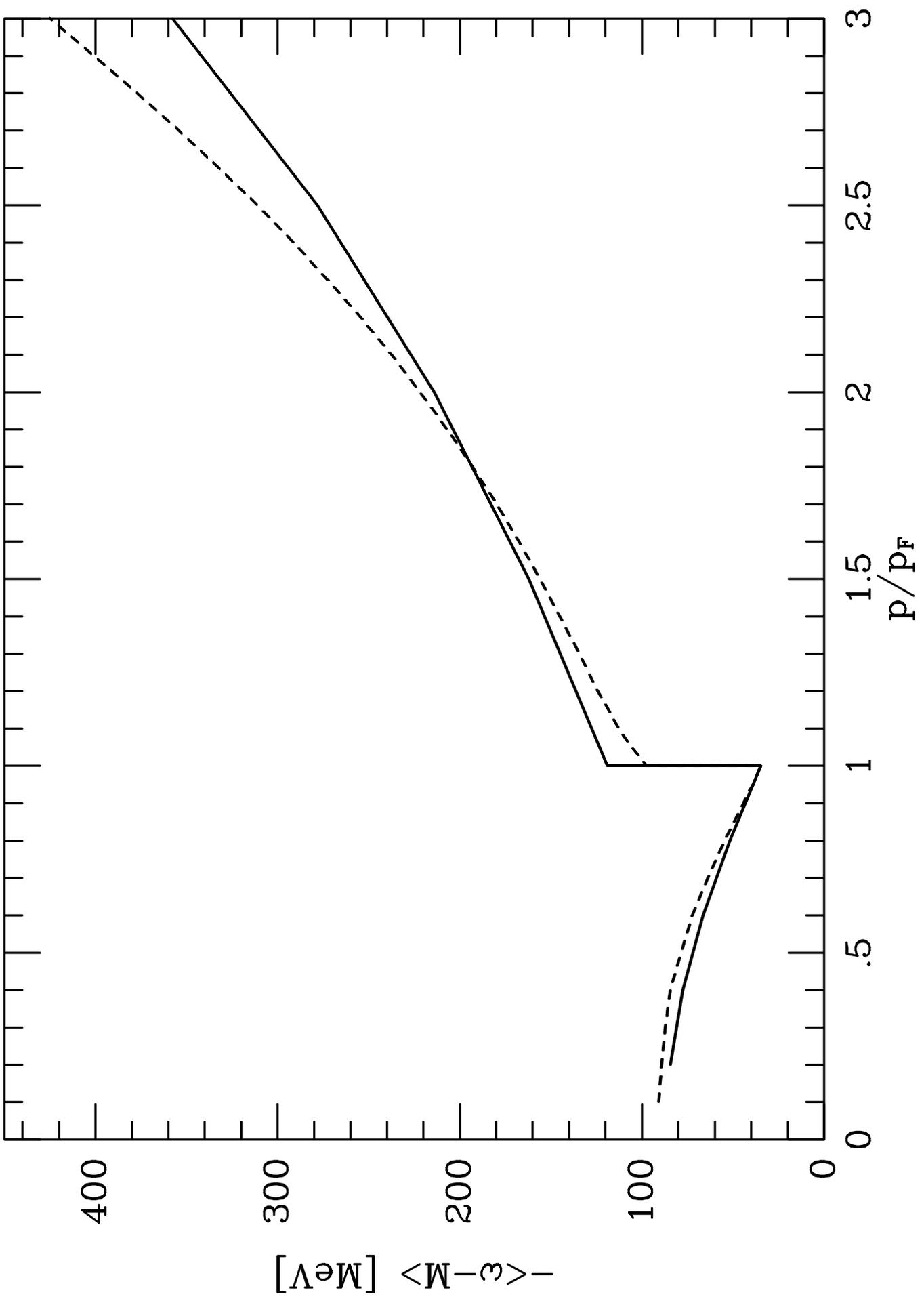
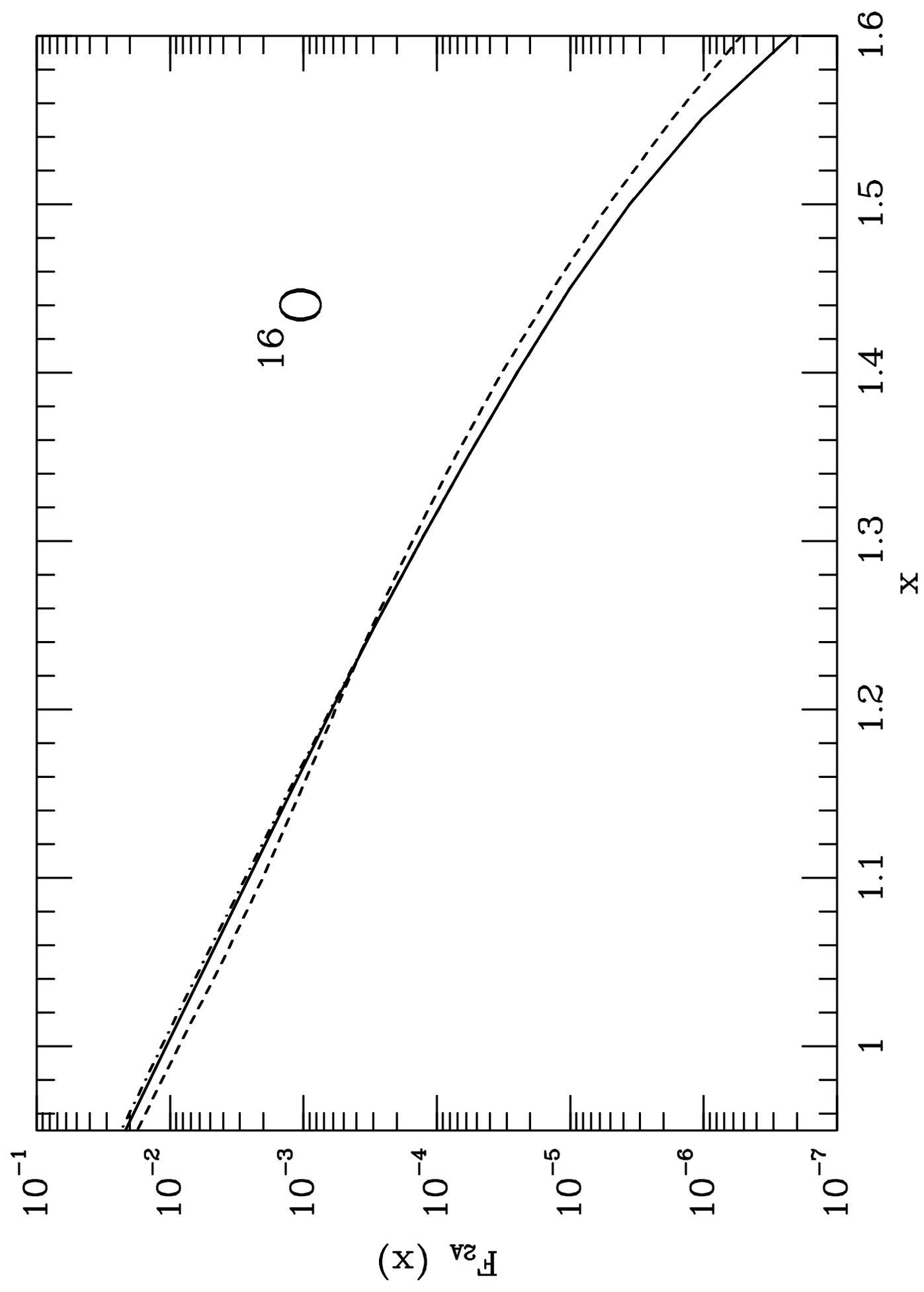
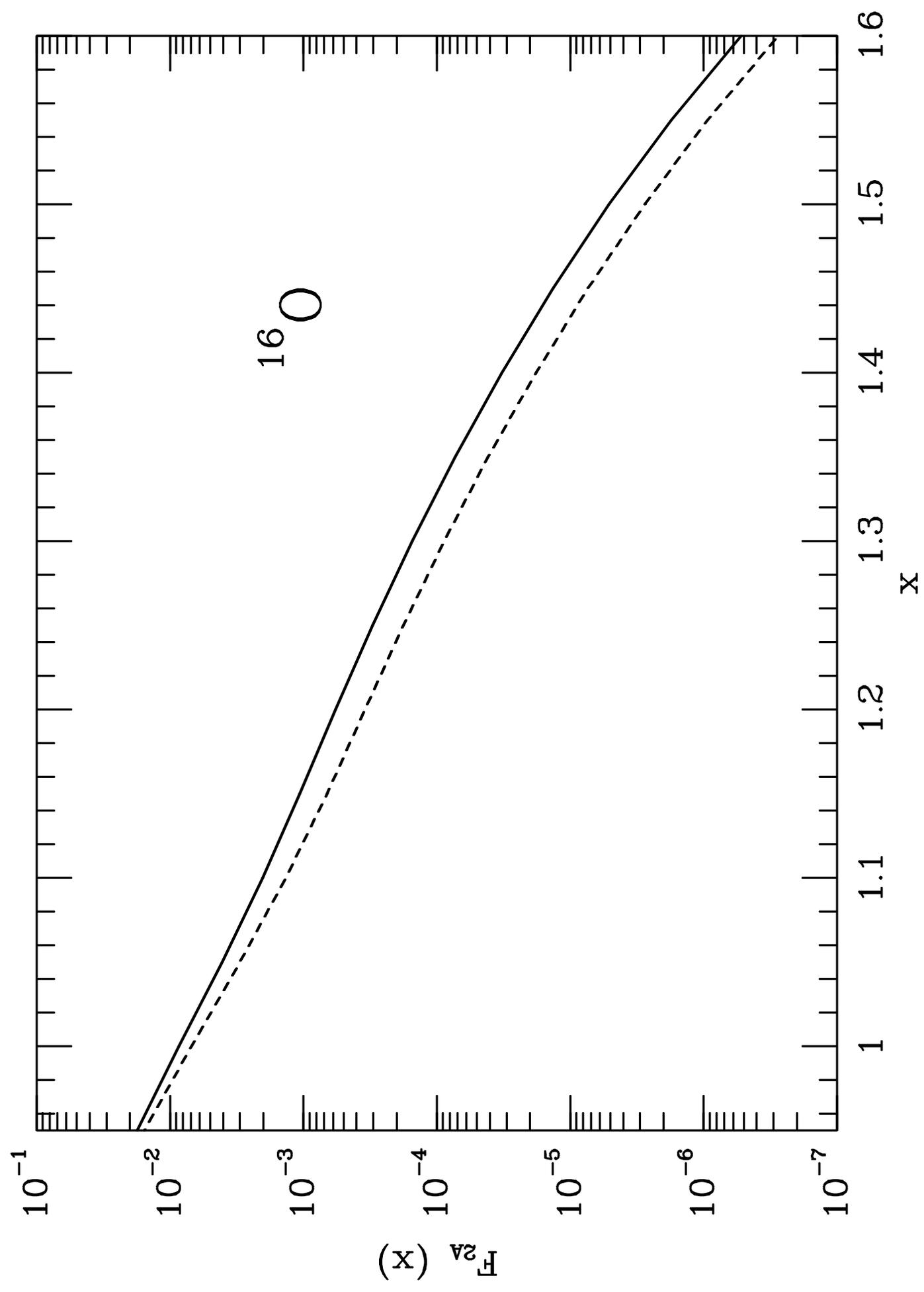
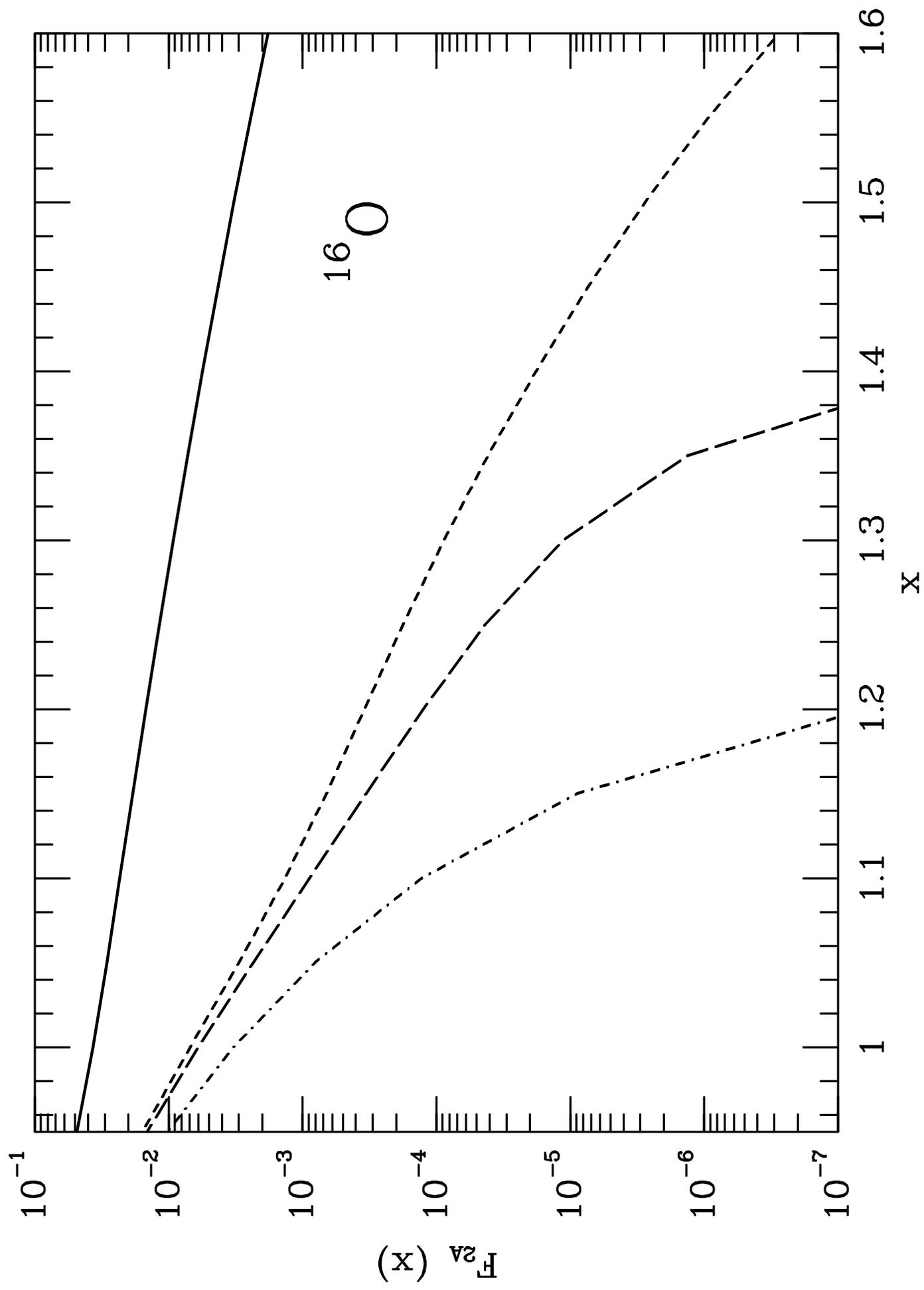
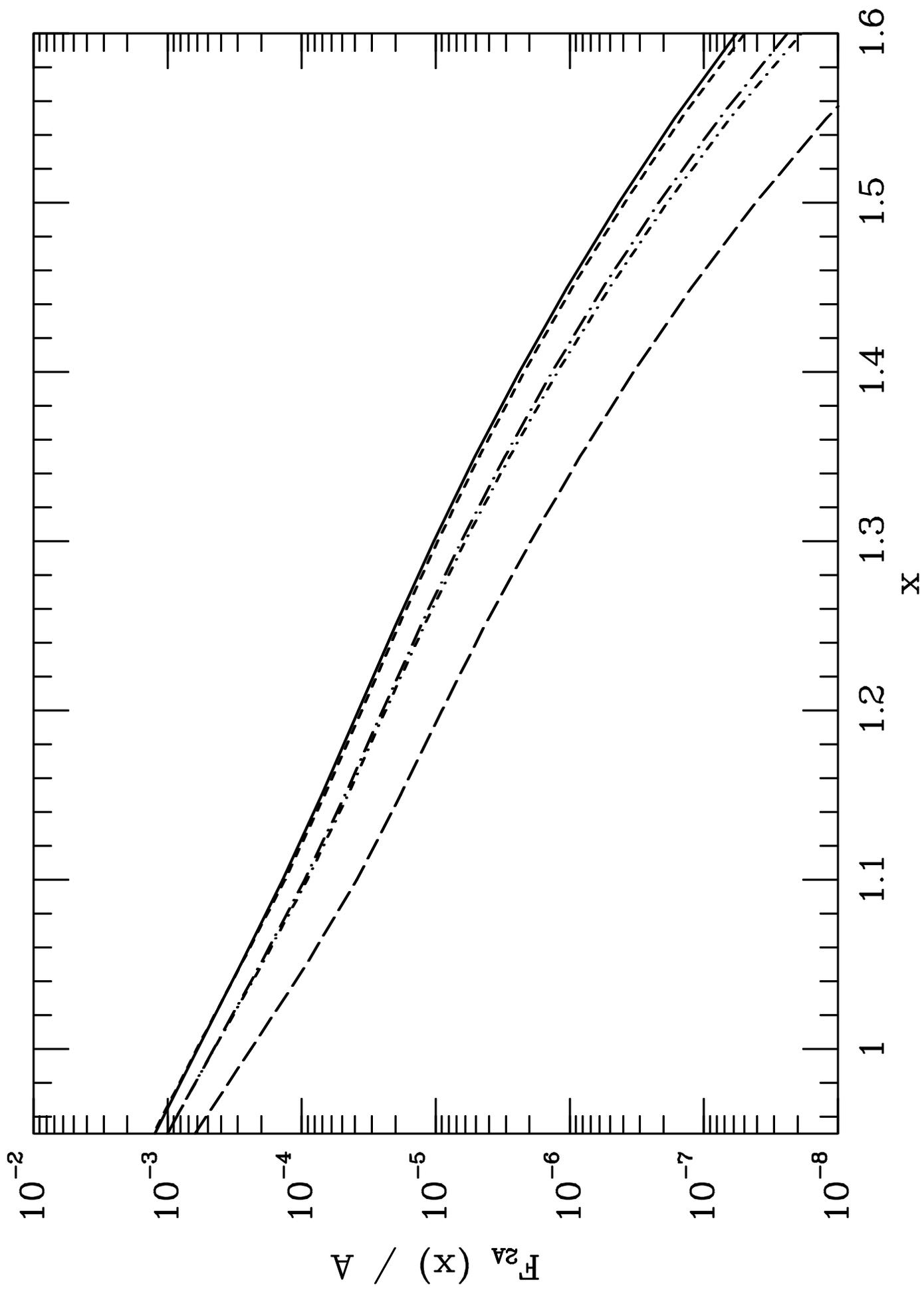
\end{document}